\documentstyle[NATO,epsf]{CRCKAPB}

\begin{opening}
\title{Photometric and Spectroscopic Monitoring of the sdBV star PG~1605+072:
}
\subtitle{The Multi-Site Spectroscopic Telescope (MSST) Project}

\author{U.\,Heber$^1$, 
S.\,Dreizler$^2$, 
S.L.\,Schuh$^2$,
S.\,O'Toole$^{1,3}$, 
C.S.\,Jeffery$^4$
}\\
\author{ 
S.\,Falter$^1$, 
V.\,Woolf$^4$,
A.\,Ahmad$^4$, 
M.\,Billeres$^5$,
S. Charpinet$^6$ 
}\\
\author{ 
O.-M.\,Cordes$^7$,
B.-Q.\,For$^8$,
E.\,Green$^8$,
E.A.\, Hyde$^8$,
A.\,Jacob$^3$
}\\
\author{ 
H.\,Kjeldsen$^9$,
S.\,Kleinman$^{10}$,
J.\,Krzesinski$^{10}$,
I.\,Lopes$^{11,12}$ 
}\\
\author{ 
S.\,Marinoni$^{13}$, 
T.\,Mauch$^3$, 
A.\,Nitta$^{10}$,
D.\,O'Donoghue$^{14}$, 
R.\,Oestensen$^{15}$, 
}\\
\author{ 
D.\,Pollacco$^{16}$,
R.\,Pereira$^{12}$,
T.\,Pereira$^{12}$,
M.D.\,Reed$^{17}$,
R.\,Silvotti$^{18}$ 
}\\
\author{ 
R.\,Townsend$^{19}$,
M.\,Vuckovic$^{20}$,
B.\,A. White$^8$,
J.\,Xiaojun$^{21}$
}\\
\institute{$^1$Dr.\,Remeis-Sternwarte,\,Univ.
Erlangen-N\"urnberg,\,Bamberg,\,D\\
$^2$Universit\"at\,T\"ubingen, D, $^3$University\,of\,Sydney,\,AUS\\
$^4$Armagh\,Observatory,\,UK, $^5$European\,Southern\,Observatory,\,Chile\\
$^6$Observatoire\,Midi-Pyrenees,\,Toulouse,\,F\\
$^7$Universit\"at\,Bonn,\,D, $^8$Steward\,Observatory,\,USA\\
$^9$University\,of\,Aarhus,\,DK, $^{10}$Apache\,Point\,Observatory, USA\\
$^{11}$Oxford\,University,\,UK, $^{12}$Instituto\,Superior\,Tecnico,
\,Lisbon,\,P\\
$^{13}$Bologna\,University,\,I, $^{14}$SAAO,\,South\,Africa\\
$^{15}$ING,\,La Palma,\,Spain, $^{16}$Queens\,University\,Belfast,\,UK\\
$^{17}$Southwest\,Missouri\,State\,University,\,USA\\
$^{18}$Osservatorio\,Astronomico\,di\,Capodimonte,\,Naples,\,I\\
$^{19}$University\,College\,London,\,UK, $^{20}$Iowa\,State\,University,\,USA
$^{21}$Beijing\,Astronomical\,Observatory,\,China}

\end{opening}

\begin{document}

\noindent
{\bf Introduction.} A small fraction of sdB stars show short-period, 
multiperiodic light 
variations and form the new class of pulsating star known as 
EC~14026 variables, after the prototype, or, alternately, as sdBV stars.
Until recently, time-series observations of pulsating sdB stars have been 
limited to photometry (for a review see Charpinet et al., \citeyear{charp01}). 
While emphasizing the great potential of asteroseismology with sdB
variables (sdBV),
Charpinet et al. also clearly demonstrate that the difficult process of mode
identification, necessarily required for an asteroseismological study, 
is even more complicated for sdBV stars than for objects
such as pulsating white dwarfs. 
While the latter tend to show a nearly
equidistant mode pattern for consecutive overtones, no such relation 
is evident for sdBV stars. Consequently, observational methods 
beyond photometry alone are needed to further progress the
interpretation of the sdBVs' pulsational behaviour.

Pulsations not only
produce photometric variations but also spectroscopic variations.
PG~1605+072 is the ideal target amongst the sdBV stars to search for such 
variations. 
It has the longest pulsation periods ($\sim$500s), the largest
amplitude variations (0.2 mag in the
optical) and the richest frequency spectrum ($>50$ modes) of all
sdBVs.  The long periods maximise the S/N that can be acquired 
within any given pulsation cycle. However, PG~1605+072 is rotating
rather rapidly at $v \sin i=39$km s$^{-1}$ \cite{heber99}, 
which makes the mode identification more
complicated due to significant rotational splitting. Therefore it
has so far been impossible to identify the modes from optical light curves.
%
%

O'Toole et al. \cite{otoole00} were the first to detect radial
velocity shifts in PG~1605+072 with an amplitude of 14
km/s (at H\,$\beta$) from time-resolved spectroscopy at medium
resolution. Time-resolved spectroscopy was subsequently 
carried out at higher spectral and temporal resolution in 2000 May
\cite{woolf02}.
Three pulsation periods were resolved by these authors and also by 
Falter et al. \cite{falter02} one year later. 
However, the amplitude ratios of the modes detected 
spectroscopically in 2000 differed from
those derived photometrically in 1998. This was also found in a
medium-resolution study by O'Toole et al. \cite{otoole02}, using
observations taken at approximately the same time as those of Woolf et al. 
Observations at higher spectral resolution obtained 
in 2001 May (Falter et al. \citeyear{falter02}) 
indicate that the amplitudes are consistent with those of the
1998 photometry again. This means that there has been a switch in
power between modes over the years. 
Falter et al. (\citeyear{falter02})
were also able to obtain multicolour photometry simultaneous with the
spectroscopy for the first time, demonstrating that the pulsation 
amplitudes in the different passbands show a wavelength dependency and
provides information on the continuum variations.
Last but not least, O'Toole et al. \cite{otoole02b} were able to measure 
Balmer line equivalent width variations which  
increase with Balmer sequence number.

The spectroscopic observations described above
all suffer from limited frequency resolution and aliasing problems
associated with finite time series or single-site observations.
Therefore an international collaboration 
has been formed to carry out observations of
sufficient length to resolve the more than 50 modes of PG~1605+072
known. The campaign has become known as the Multi-Site Spectroscopic 
Telescope (MSST).

{\bf MSST Observations.} In 2002 May and June, a coordinated series of observations of
PG~1605+072  were carried out using 15 telescopes at several
longitudes.
Because the pulsation modes of PG~1605+072 are known to be amplitude variable, 
any spectroscopic campaign has to be accompanied by photometric 
monitoring in order to measure velocity versus light variations. 
Closely spaced frequencies have to be resolved which require 
observations of several weeks with as few time gaps as possible.
A fourfold observing strategy has been developed.

\noindent 
1.) Time-resolved spectroscopy at 4-m class telescopes is used
  to  measure the  radial velocity curves with the highest accuracy
  possible. 
Due to limited 
availability of such telescopes this could only be done for a few
nights, and is insufficient to resolve all of the modes.\\
\noindent
2.) Time-resolved spectroscopy at 2-m class telescopes is used
to 
measure the radial velocity curves and the variations of equivalent widths 
for a sufficient period of time to resolve all modes.\\
\noindent
3.) Multi-band photometry is used to measure colour 
 (and hence continuum) variations for a few nights, again too short to 
 resolve all modes.\\
\noindent
4.) Broad-band photometry is used to measure light curves for a period of 
time
sufficiently long to resolve all modes. 

Table\,1 gives an overview of the data obtained for the individual
types of observations (since some of the data overlap, the
total is smaller than the sum of the contributions). 
The telescopes involved are listed below.


\parbox{10cm}{
\begin{center}
{Table 1.~}{\small\ Overall data in 2002; ff: filling factor.}\\
\vskip2mm
\small
\begin{tabular}{lrllrr}
\hline
\multicolumn{2}{l}{type of observations}& from & to & [h]&
\multicolumn{1}{c}{ff}\\
\hline
\multicolumn{2}{l}{high-resolution spectroscopy \hfill {\bf total}}
  &05/14&05/29&{\bf 27}&7\%\\[-2pt]
\cline{3-6}
\multicolumn{2}{l}{ medium-resolution spectroscopy ~~ \hfill May}
  &05/19&05/28&58&22\%\\[-2pt]
  \multicolumn{2}{r}{June}
  &06/14&06/26&94&30\%\\[-2pt]
\cline{3-6}
\multicolumn{2}{l}{photometry \hfill {\bf total}}
&05/05&06/04&{\bf 328}&44\%\\[-2pt]
          &WET&05/05&05/21&127&31\%\\[-2pt]
          &MSST&05/14&06/04&272&51\%\\[-2pt]
          &multi-colour&05/19&05/22&21&22\%\\[2pt]
\hline
\end{tabular}
\label{tab:Schuh1_t1}
\end{center}

\normalsize
}

\noindent
1.) Time-resolved spectroscopy of spectral resolution
$\sim$1\AA\ was carried out during
eight nights at the following telescopes:
the Calar Alto  3.5-m (Spain), 
the European Southern Observatory 3.5-m NTT (Chile) and the 
Apache Point 3.5-m (New Mexico, USA). 
Spectra were taken by trailing the star along the slit
(Falter et al. 2002)
with a time resolution of typically 30\,s. 
The data set corresponds to about 4000 individual spectra.\\
\noindent 
2.) Time resolved spectroscopy of resolution $\sim$3\AA\ was
carried out during 40 nights at the following telescopes,
the Mount Stromlo and Siding Spring Observatory 2.3-m
(Australia), the Nordic Optical 2.56-m  (La Palma, Spain), 
the European Southern Observatory 1.5-m Danish (La Silla, Chile),
and the Steward Observatory 2.3-m Bok (Arizona, USA). 
About 10\,000 spectra have been secured.\\
\noindent 
3.) PG~1605+072 was observed in four photometric bands with the BUSCA 
instrument (the instrumental ``U'',``B'',``R'',``I'' passbands are used)
at the Calar Alto 2.2-m telescope for three nights.\\
\noindent 
4.) A combination of BG\,39 filter, B-band and white light photometry 
was carried out  at the following telescopes:
the Mount Stromlo and Siding Spring Observatory 1-m (Australia), 
the Beijing Astronomical Observatory 0.85-m (China), 
the South African Astronomical Observatory 1-m (South Africa), 
the Loiano 1.5-m (Italy),
the Jacobus Kapteyn 1-m  (La Palma, Spain),
the Fick Observatory 0.6-m (Iowa, USA)
and the McDonald Observatory 2.1-m (Texas, USA).
Data were acquired covering more than 272 hours.
In addition, the Whole Earth Telescope observed 
PG~1605+072 in 2002 May for 127 hours \cite{schuh02}.


\vspace*{0.2cm}
\noindent
{\bf Summary and Conclusion.}
The Multi-Site Spectroscopic Telescope 
is a virtual instrument and is also the name
of a collaboration that aims to open up a new observational window  
to provide access to a 
mode identification for and an asteroseismological analysis of the pulsating 
sdB star PG~1605+072.
Although the primary aim is to obtain time resolved spectroscopy it also 
includes the most extended photometric monitoring campaign for PG~1605+072.   

The data gathered during the 
MSST campaign constitutes a huge improvement over 
earlier efforts. Its usefulness however depends in exactly the same
way on the availability of good models for an interpretation of the
data. 
The time dependent surface 
distribution of $\mathrm{T}_{eff}$, ${\mathrm{log}}\,g$ and 
the velocity field will be
determined using Townsend's \cite{townsend97} program BRUCE
which also takes effects caused
by rotation into account. 
Grids of synthetic spectra calculated from LTE and NLTE model atmospheres
will be used to carry out the surface 
integration to derive phase-dependent flux spectra with a code 
developed by Falter \cite{falter01}. 
Furthermore, detailed pulsation models 
(Charpinet et al., \citeyear{charp01}) are at hand for 
the comparison with 
the observational frequency pattern.

\vspace*{0.2cm}
\noindent
{\bf ACKNOWLEDGEMENTS.} We thank the various time allocation committees who 
awarded such a large amount of observing time. The Whole Earth Telescope 
collaboration is thanked for adopting PG~1605+072 as an alternate target 
during XCOV22. Finally we gratefully acknowledge
the support of staff at the observatories involved in the campaign.   



\end{document}